\documentclass[prd,amsmath,showpacs,preprintnumbers,superscriptaddress,
               floatfix,twocolumn,h-physrev4]{revtex4}
\usepackage{times}
\usepackage{graphicx}
\usepackage{amssymb}
\usepackage{mathrsfs}
\usepackage{slashed}
\usepackage{xcolor}
\usepackage[normalem]{ulem}
\newcommand{\ro}{\mathrm}
\newcommand{\ca}{\mathcal}

\DeclareMathOperator{\SU}{SU}

\begin{document}

\title{Finite-volume and partial quenching effects in the magnetic polarizability of the neutron}

\author{J. M. M. Hall} 
\affiliation{Special Research Centre for the Subatomic Structure of
  Matter (CSSM), School of Chemistry and Physics, University of Adelaide, 
  Adelaide, South Australia 5005, Australia}

\author{D. B. Leinweber}
\affiliation{Special Research Centre for the Subatomic Structure of
  Matter (CSSM), School of Chemistry and Physics, University of Adelaide, 
  Adelaide, South Australia 5005, Australia}

\author{R. D. Young} 
\affiliation{Special Research Centre for the Subatomic Structure of
  Matter (CSSM), School of Chemistry and Physics, University of Adelaide, 
  Adelaide, South Australia 5005, Australia}
\affiliation{ARC Centre of Excellence for Particle Physics at the Terascale, 
School of Chemistry and Physics, University of Adelaide, 
  Adelaide, South Australia 5005, Australia}

\begin{abstract}
There has been much progress in the experimental measurement of the
electric and magnetic polarizabilities of the nucleon.  Similarly,
lattice QCD simulations have recently produced dynamical QCD results
for the magnetic polarizability of the neutron approaching the chiral
regime.  In order to compare the lattice simulations with experiment,
calculation of partial quenching and finite-volume effects is required
prior to an extrapolation in quark mass to the physical point.  These
dependencies are described using chiral effective field theory.
Corrections to the partial quenching effects associated with the
sea-quark-loop electric charges are estimated by modelling corrections 
to the pion cloud. These are compared to the
uncorrected lattice results.  In addition, the behaviour of the
finite-volume corrections as a function of pion mass is explored.  Box
sizes of approximately $7$ fm are required to achieve a result within
$5\%$ of the infinite-volume result at the physical pion mass.  A
variety of extrapolations are shown at different box sizes, providing
a benchmark to guide future lattice QCD calculations of the magnetic
polarizabilities.  A relatively precise value for the physical
magnetic polarizability of the neutron is presented, $\beta^n =
1.93(11)^{\ro{stat}}(8)^{\ro{sys}} \times 10^{-4}\,\ro{fm}^3$, which
is in agreement with current experimental results.
\end{abstract}

\maketitle
\preprint{ADP-13-37/T857}

\section{Introduction}

The study of the electric and magnetic polarizabilities of the nucleon
is a topic of intense ongoing interest.  Although measurement of the
sum of the polarizabilities from Compton scattering has been achieved
experimentally for some time \cite{Bernabeu:1974zu,Bernard:1991rq}, a
direct determination of the individual electric and magnetic
polarizabilities still presents a challenge, with uncertainties
remaining large
\cite{Kossert:2002jc,Kossert:2002ws,McGovern:2012ew,Griesshammer:2012we}. In
the case of the neutron, recent values of $\beta^n$ include $4.1\pm
2.0$ \cite{Griesshammer:2012we}, 
$3.7\pm 2.0$ \cite{Beringer:1900zz} 
and $2.7^{+2.2}_{-2.4}$
\cite{Kossert:2002jc,Kossert:2002ws}, in units of
$10^{-4}\,\ro{fm}^3$.

Recent improvements in lattice QCD techniques in the treatment of
Landau levels \cite{Tiburzi:2012ks} and the simulation of uniform
magnetic fields with improved boundary conditions
\cite{Primer:2012pv,Primer:2013pva} offer new insights into the
polarizabilities of the nucleon.  When comparing lattice results with
experiment, care must be taken in extrapolating the results to the
chiral limit while incorporating finite-volume effects.  The latter
have been shown to be significant even at modern lattice volumes
\cite{Hall:2012pk,Hall:2012yx,Hall:2013oga,Greil:2011aa,Albaladejo:2012jr,Doring:2013glu,Briceno:2013lba}.

Chiral perturbation theory ($\chi$PT) represents an important tool for
performing chiral extrapolations of lattice results to the physical
point.  Though lattice simulations are now approaching the chiral
regime
\cite{Boinepalli:2006xd,Aoki:2008sm,Kenway:2008wia,Collins:2011mk,Bouchard:2013eph,Prelovsek:2013ela,Aoki:2013ldr,Inoue:2013nfe},
multiple pion-mass points must be used to constrain the parameters of
the extrapolation.  These data sets typically extend outside the
chiral power-counting regime of $\chi$PT.  It has been demonstrated
that use of $\chi$PT outside its region of applicability leads to
incorrect results \cite{Leinweber:2003dg,Leinweber:2005xz}.
Therefore, an extension of chiral effective field theory ($\chi$EFT)
with improved convergence properties will be used.  The approach
incorporates a resummation of the higher-order terms of the chiral
expansion, while being exactly equivalent to $\chi$PT within the
power-counting regime
\cite{Leinweber:2003dg,Leinweber:2005xz}.  The size of
the total contribution from the higher-order terms is determined by a
finite-valued energy scale which has been linked to the intrinsic
scale associated with the size of the hadron under investigation
\cite{Hall:2010ai,Hall:2011en}.

In this paper, the focus is on the magnetic polarizability of the neutron,
and connect recent lattice QCD results from the CSSM
\cite{Primer:2012pv,Primer:2013pva} to contemporary experimental
results, providing a sound comparison of theory and experiment.  The
lattice QCD results are founded on the PACS-CS configurations
\cite{Aoki:2008sm} made available via the International Lattice Data
Grid (ILDG) \cite{Beckett:2009cb}.  These dynamical QCD results from
the simulation are analyzed using $\chi$EFT.  
A particular difficulty, addressed in this paper, is that the best
lattice QCD results have yet to incorporate the contributions from
photon couplings to sea-quark loops comprising the meson dressings of
$\chi$EFT.  Our choice of regularization scheme facilitates the 
 modelling of 
the corrections \cite{Detmold:2006vu} associated with these effects.

In the following sections, the methods of unquenching, finite-volume
corrections, and chiral extrapolations are established, and a
prediction for the magnetic polarizability of the neutron is reported.  
This prediction is complemented by a variety of finite-volume
extrapolations at different box sizes, providing a benchmark to guide
future lattice QCD calculations of the magnetic polarizabilities.

\section{Lattice QCD}

In calculating the magnetic polarizability in lattice QCD, a
background magnetic field $B$ is introduced on the lattice by
multiplying each gauge link by a certain phase factor
\cite{Bernard:1982yu,Smit:1986fn,Burkardt:1996vb,Lee:2005ds,Lee:2005dq,Lee:2008qf,Primer:2012pv,Primer:2013pva}.
In the weak-field limit, the resultant energy shift of the nucleon is
dependent on the magnetic moment $\vec{\mu}$, and the magnetic
polarizability $\beta$, through
\begin{equation}
E(B) = M_N - \vec{\mu}\cdot\vec{B} + \frac{e|B|}{2M_N} - 2\pi \beta\,B^2 + \ca{O}(B^3). 
\end{equation}

The period boundary conditions restrict the possible values of the
magnetic field strength, based on the number of lattice sites $N_x$
and $N_y$ in the $x$ and $y$ directions, leading to the quantization
condition
\begin{equation}
qBa^2 = \frac{2\pi n}{N_x N_y}, \quad n\in \mathbb{Z}, 
\end{equation}
for a quark charge $q$. 

\begin{figure}
\includegraphics[height=0.95\hsize,angle=90]{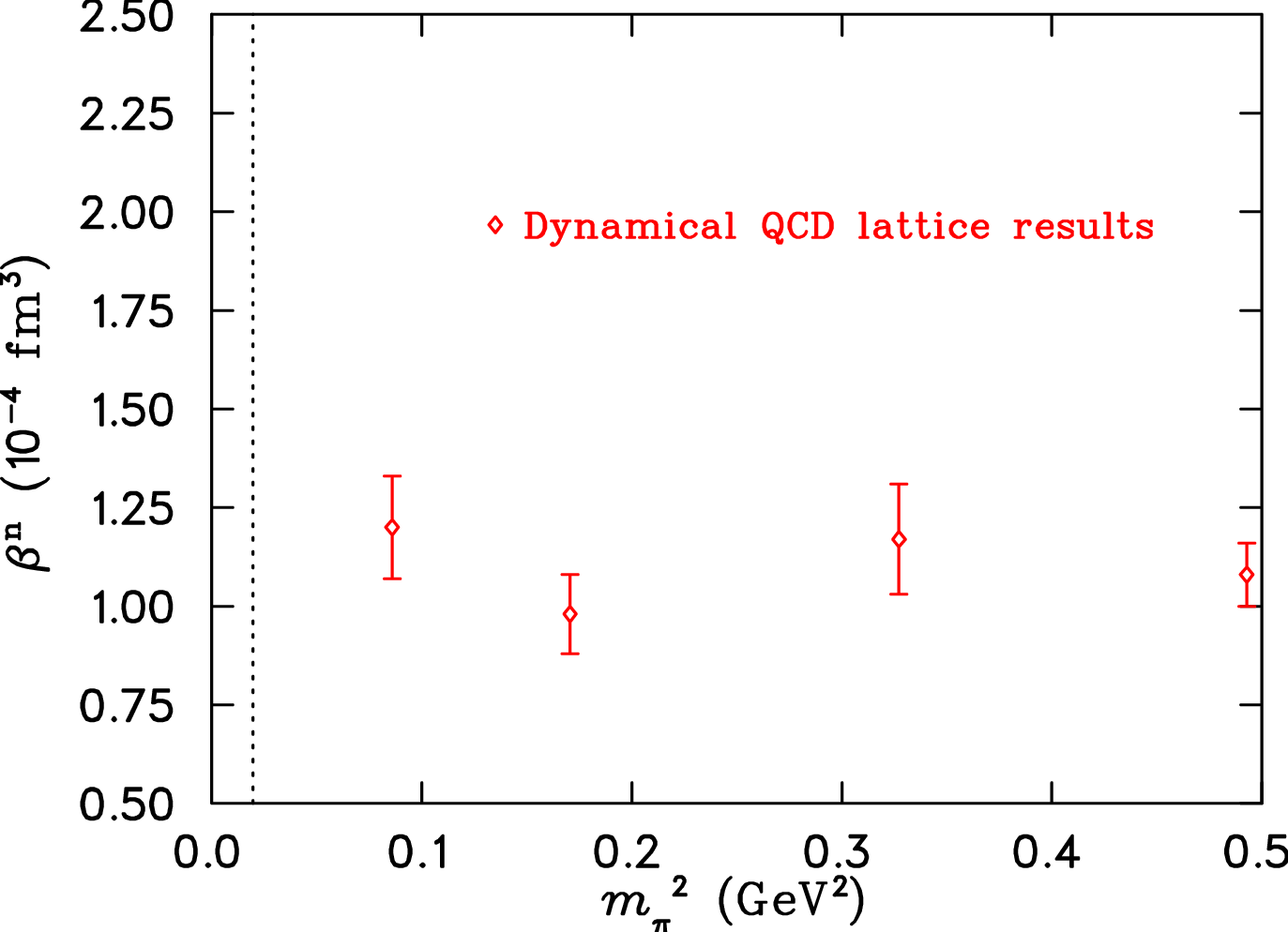}
\vspace{-11pt}
\caption{\footnotesize{(color online). The magnetic polarizability of
    the neutron $\beta^n$, from $2+1$ flavour lattice QCD simulations
    by the CSSM \cite{Primer:2012pv,Primer:2013pva}.  The results are
    based on the PACS-CS configurations \cite{Aoki:2008sm} available
    via the ILDG \cite{Beckett:2009cb}.  }}
\label{fig:data}
\end{figure}

The background field method is applied to the PACS-CS configurations
\cite{Aoki:2008sm} obtained via the ILDG \cite{Beckett:2009cb}, which
use the $2+1$ flavour improved clover fermion action and the Iwasaki
gauge action.  The lattice results used in this analysis are presented
in Fig.~\ref{fig:data}, utilizing the Sommer scale parameter
\cite{Sommer:1993ce}, $r_0 = 0.49$ fm
\cite{Primer:2012pv,Primer:2013pva}. 
Note that all the lattice points considered satisfy $m_\pi L > 4.45$  
such that the use of finite-volume $\chi$EFT in analysing these results 
is appropriate.

In computing the polarizabilities, contributions from photon couplings
to disconnected sea-quark loops have not yet been included.  These
need to be accounted for prior to making 
a comparison with experimental results.
In the case of the neutron, partially quenched $\chi$EFT is
used to determine the appropriate chiral behaviour of the
polarizability in both the partially quenched scenario of the lattice
results and full QCD.

Other calculations of the magnetic polarizability of the neutron 
\cite{Lee:2005ds,Lee:2008qf} use a different approach. While 
the linearization of the U$(1)$ field breaks gauge invariance 
\cite{Primer:2013pva}, the main concern in this alternative data set 
is the use of the Dirichlet 
boundary condition in a spatial direction breaking the symmetry of 
the $3$-torus, which may give rise to significant systematic errors 
due to artefacts at the boundary. 
Since finite-volume 
$\chi$EFT employs periodic boundary conditions in all spatial directions, 
these lattice results are not compatible with the formalism, and are 
therefore not used in this investigation.

\section{Chiral effective field theory}
\label{sec:eft}

%
%
The electric and magnetic polarizabilities $\alpha$ and $\beta$,
respectively, may be defined in terms of two independent parameters, 
($A$, $B$),   
obtained from expanding 
the Compton tensor \cite{Bernabeu:1974zu}, 
\begin{equation}
T_{\mu\nu}(p,q) = \int\!\ro{d}^4x\,e^{i q\cdot x}\,\langle N(p)|\ca{T}\left\{J_{\mu}^{\ro{em}}(x)J_{\nu}^{\ro{em}}(0)\right\}|N(p)\rangle.
\end{equation}
 They are defined as
\begin{equation}
\alpha + \beta = -\frac{e^2 m}{2\pi}\frac{\partial^2 A(s)}{\partial s^2}\Big|_{s=m^2}, \quad
\beta = -\frac{e^2}{4\pi m}B(s=m^2).
\label{eqn:beta}
\end{equation}

The interaction vertices are sourced from the relevant first-order
interaction Lagrangian of heavy-baryon $\chi$PT, which includes the $\Delta$ baryon
transitions \cite{Wang:2007iw,Wang:2008vb} 
\begin{align}
\label{eqn:lag}
\ca{L}_{\ro{HB}\chi\ro{PT}}^{(1),\ro{int}} &= i\frac{g_A}{2f_\pi}\bar{\Psi}_v
\gamma_5 \sigma_{\mu\nu}v^\nu {\tau^a}\partial^\mu {\pi^a}\Psi_v \nonumber\\
&+i\frac{\ca{C}}{2 f_\pi} 
\bar{\Psi}_v T^a \sigma_{\mu\nu}\Delta^\nu_v \partial^\mu \pi^a + \ro{h.c.}
\end{align}
$T^a$ are the isospin doublet-quartet transition matrices.  The octet
couplings $D$ and $F$ are derived from the experimental value of
$g_A$, obeying the condition $D+F = g_A$ and the $\SU(6)$ symmetry
relation $F = \frac{2}{3}D$.

\begin{figure}[tp]
\centering
\includegraphics[height=80pt]{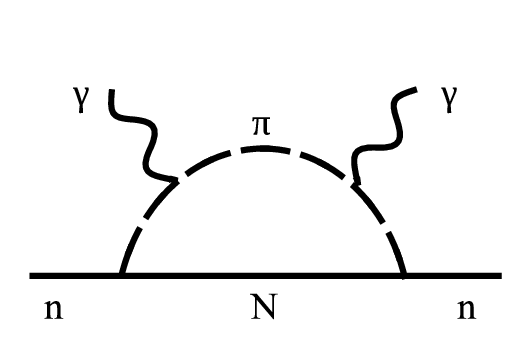}
\caption{\footnotesize{The leading-order pion loop contribution to the magnetic polarizability of the neutron.}}
\label{fig:polSEa}
\centering
\includegraphics[height=80pt,angle=0]{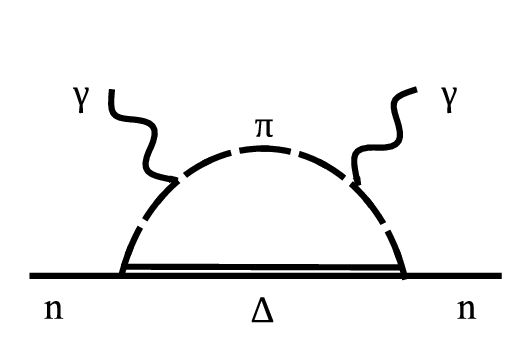}
\caption{\footnotesize{The leading-order pion loop contribution to the
    magnetic polarizability of the neutron, allowing transitions to
    nearby and strongly coupled $\Delta$ baryons.}}
\label{fig:polSEb}
\end{figure}

The calculation of the leading-order loop contributions follows
Refs.~\cite{Detmold:2006vu,Lensky:2009uv} and uses the heavy-baryon
approximation.  In the finite-range regularization approach, the
leading nonanalytic contribution to the chiral expansion comes
entirely from the diagram shown in Fig.~\ref{fig:polSEa}, for the
above parametrization of $\beta^n$ (contributing to $A_2$ from
Ref.~\cite{Griesshammer:2012we}).
Transitions to a $\Delta$ baryon at leading order 
are shown in Fig.~\ref{fig:polSEb}. For a finite mass-splitting 
$\Delta$, Fig.~\ref{fig:polSEb} contributes a log term rather than $1/m_\pi$ \cite{Lensky:2009uv}. 

By performing the pole integration over $k_0$ and taking the forward
scattering limit ($q\cdot q' \rightarrow 0$), one obtains a
three-dimensional integral form that can easily be adapted to estimate
finite-volume corrections
\cite{Armour:2005mk,Hall:2010ai,Hall:2013qba}
\begin{align}
\label{eqn:loopa}
\beta^{(\pi N)}(m_\pi^2) &= \frac{e^2}{4\pi}\frac{1}{288\pi^3f_\pi^2}\,
\chi_N\!\int\!\ro{d}^3k\,\frac{\vec{k}^2}{(\vec{k}^2+m_\pi^2)^3}, \\
\beta^{(\pi\Delta)}(m_\pi^2) &= \frac{e^2}{4\pi}\frac{1}{288\pi^3f_\pi^2}\, 
\chi_\Delta\!\int\!\ro{d}^3k\,\times\nonumber \\
&
\frac{\omega_{\vec{k}}^2\Delta(3\omega_{\vec{k}}+\Delta) + \vec{k}^2(8\omega_{\vec{k}}^2+9\omega_{\vec{k}}\Delta+3\Delta^2)}
{8\omega_{\vec{k}}^5(\omega_{\vec{k}}+\Delta)^3},
\label{eqn:loopb}
\end{align}
where $\omega_{\vec{k}} = \sqrt{\vec{k}^2+m_\pi^2}$ is the energy
carried by the pion with three-momentum $\vec{k}$, $\Delta \equiv
M_\Delta-M_N = 292$ MeV, and the pion decay constant is taken to be
$f_\pi=92.4$ MeV.
The standard coefficients for full QCD are given by   
\begin{align}
\chi_N &= 2 g_A^2,\\
 \chi_\Delta&=\frac{16}{9}\ca{C}^2,
\end{align}
with the coupling constants taking the values $g_A=1.267$,
$\ca{C}=-1.52$.  Modifications to the couplings to accommodate partial
quenching effects are explained in Sec.~\ref{subsec:unQ}.

The leading-order $1/m_\pi$ contribution to the magnetic
polarizability of the neutron has the established coefficient
\cite{Bernard:1991ru,Bernard:1993bg,Bernard:1993ry}
\begin{equation}
\beta^{\pi N}(m_\pi^2) = \frac{e^2g_A^2}{768\pi^2 f_\pi^2}
\frac{1}{m_\pi} \equiv \chi'\frac{1}{m_\pi}. 
\end{equation}
The chiral expansion of the magnetic polarizability of the 
neutron at this order is
\begin{equation}
\label{eqn:expsn}
\beta^n 
= \frac{\chi'}{m_\pi} + c_0 - \frac{16}{9\Delta}\chi'\log(m_\pi/\mu) +\ca{O}(m_\pi),
\end{equation}
where $\mu$ is an arbitrary mass scale associated with the logarithm, henceforth set to $1$ GeV.

Finite-volume effects are evaluated following the prescription
described in Refs.~\cite{Hall:2010ai,Hall:2011en,Hall:2012pk}.  These
are estimated by evaluating the corrections associated with replacing
the continuum integrals of Eqs.~(\ref{eqn:loopa}) and
(\ref{eqn:loopb}) with finite sums over the momenta available on the
lattice.  Using this method, the finite-volume corrections are stable
for large values of the regularization scale, and are numerically
equivalent to the algebraic approach described in
Refs.~\cite{AliKhan:2003cu,Beane:2004tw}.

When applying $\chi$EFT to lattice QCD results, pion-mass values
extending outside the chiral power-counting regime are typically
considered, and use of standard $\chi$PT in this region inevitably
leads to a badly divergent chiral series
\cite{Leinweber:2003dg,Leinweber:2005xz}.  The identification of an
intrinsic scale for the regularization of loop integrals provides a
robust method for resumming the higher-order terms of the chiral
expansion and calculating the low-energy coefficients of $\chi$PT
\cite{Hall:2010ai,Hall:2011en,Hall:2012pk,Hall:2013oga}.  

In evaluating the loop integrals of the effective field theory, a
dipole regulator, $u^2(k,\Lambda) = (1 + \vec k^2/\Lambda^2 )^{-4}$, 
is introduced into the
integrands to ensure only soft momenta flow through the effective
field theory degrees of freedom.  Through an examination of the flow
of the low-energy coefficients constrained by the lattice QCD results
as the regulator parameter, $\Lambda$ is varied, and with an
understanding of the dependence of this flow on the range of pion
masses considered in the chiral expansion, one can identify a value of
$\Lambda$ which provides low-energy coefficients independent of the
pion mass range considered, and are therefore consistent with the
low-energy coefficients of $\chi$PT in the PCR
\cite{Hall:2010ai,Hall:2011en,Hall:2012pk,Hall:2012iw,Hall:2013oga}.

%
A weighted average across studies of the leading-order chiral
coefficients of the nucleon mass \cite{Hall:2010ai,Hall:2011jr} (using
PACS-CS results \cite{Aoki:2008sm}), magnetic moment
\cite{Hall:2012pk} and the electric charge radius \cite{Hall:2013oga}
(using QCDSF results \cite{Collins:2011mk}) leads to a regulator
parameter of $\bar{\Lambda}^{\ro{scale}} = 0.99(27)$ GeV.

For the present case the dipole regulator is introduced into the
integrands of Eqs.~(\ref{eqn:loopa}) and (\ref{eqn:loopb}).  In light
of our additional task of correcting for the partial quenching of the
lattice QCD simulations, the value $\Lambda = 0.80$ GeV is adopted 
\cite{Young:2002cj,
Leinweber:2004tc,
Leinweber:2006ug,
Wang:1900ta,
Wang:2008vb,
Wang:2013cfp
}, 
in agreement with the intrinsic scale identified in previous studies
\cite{Hall:2010ai,Hall:2011jr,Hall:2012pk,Hall:2013oga}.
This particular regulator mass has been shown to define a pion cloud
contribution to masses \cite{Young:2002cj}, magnetic moments
\cite{Leinweber:2004tc} and charge radii \cite{Wang:2008vb}, which 
enables one to model the correction to the pion cloud encountered in
unquenching and to reproduce experimental measurements. 
As explained in Refs~\cite{Young:2002cj,
Leinweber:2004tc,
Leinweber:2006ug,
Wang:1900ta,
Wang:2008vb,
Wang:2013cfp
}, this particular choice of regulator parameter defines a neutron core 
contribution, which does not differ significantly
 between partially quenched QCD 
and full $2+1$ flavour QCD. In making this connection, one makes 
the model assumption that the core is insensitive to
 sea-quark loop contributions.

\subsection{Partially quenched chiral effective field theory}
\label{subsec:unQ}

In this section, an independent calculation is presented for the
calculation of the loop coefficients for partially quenched $\chi$PT.  The
approach is complementary to the graded-symmetry approach selected by
Detmold \textit{et al.} \cite{Detmold:2006vu} and provides an
alternative picture of the process based on standard nuclear physics
processes.  In addition, the unquenching procedure is outlined, 
which corrects the lattice simulation results to full QCD.

The procedure for obtaining the partially quenched $\chi$PT
coefficients of nonanalytic terms follows that described in
Ref.~\cite{Leinweber:2002qb}.  First, one separates the contribution
from each quark-flow diagram into ``valence-valence'', ``valence-sea'' and
``sea-sea'' contributions. These describe whether the two photons couple
to valence or sea quarks in the intermediate states available in
Figs.~\ref{fig:polSEa} and \ref{fig:polSEb}.  

\begin{figure}[tp]
\centering
\includegraphics[width=0.7\hsize]{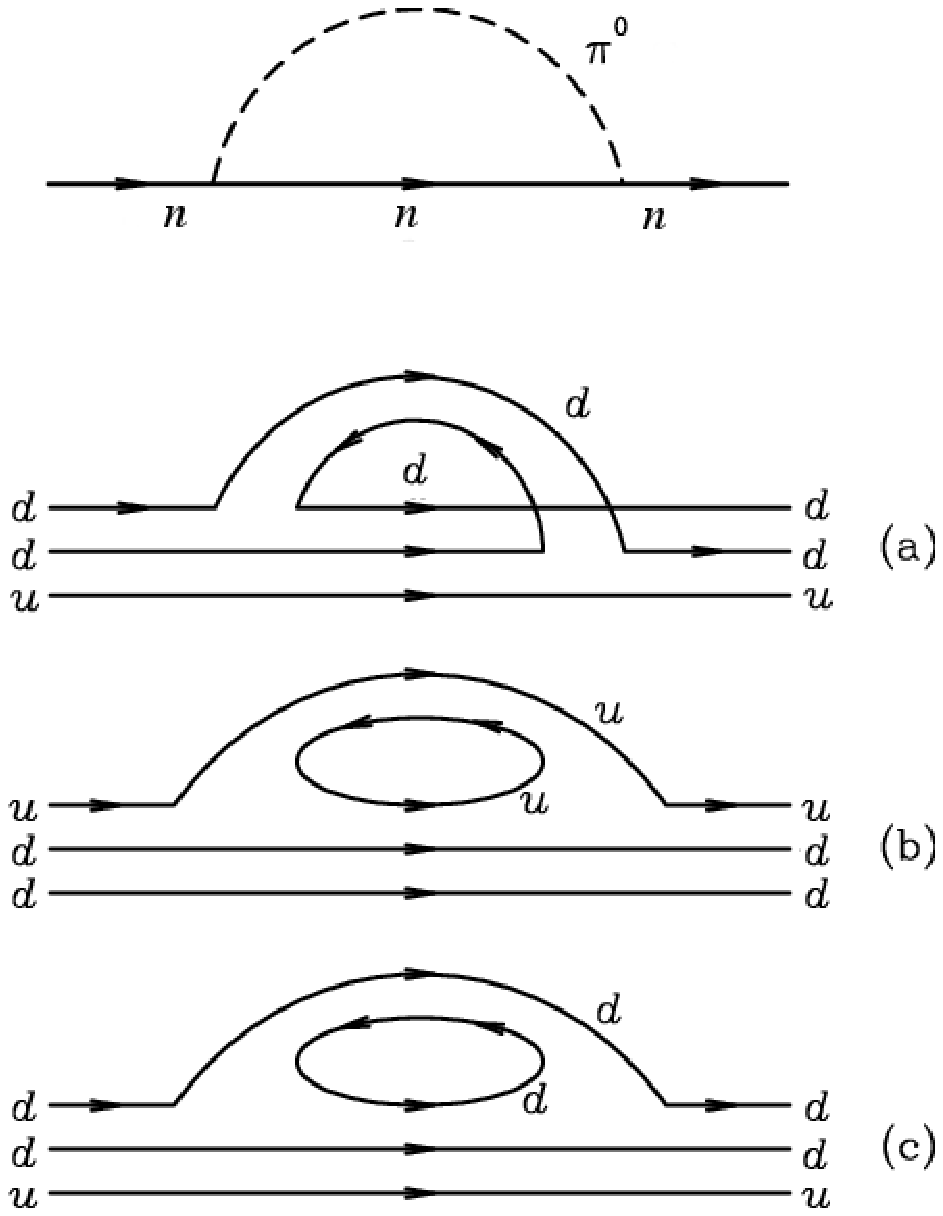}
\caption{\footnotesize{Example: the decomposition of the process
    $n\rightarrow n \pi^0$ into its possible one-loop quark-flow
    diagrams.  The configuration of photon couplings to valence and sea
    quarks will determine the coefficients of partially quenched
    $\chi$PT.}}
\label{fig:nnpi0}
\end{figure}

Starting with $n\rightarrow N\pi$, and selecting the $n\rightarrow n
\pi^0$ channel, one writes out all the possible quark-flow diagrams,
prior to attaching external photons to the meson, as shown in
Fig.~\ref{fig:nnpi0}. Diagram~4(a) contains only valence quarks, and
therefore can only contribute to the valence-valence
sector. Diagrams~4(b) and (c) may contribute to all three sectors,
as one or both photons may be attached to the valence- or sea-quark
lines of the intermediate meson. These occur in proportion to the
quark charges.  

While the neutrality of the $\pi^0$ meson ensures the total leading-order 
contribution is zero, this occurs through a combination of
valence-valence, valence-sea, and sea-sea contributions, with the latter two
omitted in the lattice QCD simulations.  For example, in the case of
diagram 4(b), the coupling of photons to either valence- or
sea-quark lines generates
\begin{align}
\label{eqn:nb}
\chi_{nn\pi^0}^{diag (b)} &\propto (q_u^2 + 2\, q_u\, q_{\bar{u}} +
q_{\bar{u}}^2) \, ,
\\ 
&\propto (q_u + q_{\bar{u}})^2 = 0 \, . 
\end{align} 
The three terms in Eq.~(\ref{eqn:nb}) contribute to valence-valence,
valence-sea and sea-sea, respectively. For quark charges $q_u = +2/3$,
$q_{\bar{u}} = -2/3$, clearly the total contribution from
$n\rightarrow n\pi^0$ vanishes, as expected.  

Setting electric charges aside, the $SU(3)$ flavour coupling for
diagram 4(b) alone is obtained by temporarily replacing the up-quark
sea-quark-loop with a strange-quark \cite{Leinweber:2002qb}.  This
correctly isolates the quark flow diagram only containing a disconnected
sea-quark-loop flow: 
\begin{equation}
\label{eqn:gc8}
\chi_{nn\pi^0}^{diag (b)} \propto \chi_{K^+\Sigma^-}^2 =  2(D-F)^2 \, . 
\end{equation}

\begin{table*}[tp]
 \caption{\footnotesize{The relative contributions to the
     leading-order loop integrals of Figs.~\ref{fig:polSEa} and
     \ref{fig:polSEb}.  The numerical value of the couplings can be
     obtained by inserting the appropriate quark charges, and noting
     $\chi_{K^+\Sigma^-}^2 = 2(D-F)^2$,\quad $\chi^2_{K^0\Sigma^0}=
     (D-F)^2$ and $\chi^2_{K^0\Lambda}= (D+3F)^2/3$ for the octet
     intermediate states and $\chi_{K^+\Sigma^{*-}}^2 = 4\, \ca{C}^2 /
     9$ and $\chi_{K^0\Sigma^{*0}}^2 = 2\, \ca{C}^2 / 9$ for decuplet
     intermediate states.  The valence-valence sector can be
     calculated by subtracting the two other sectors from the total
     result.}}  \newcommand\T{\rule{0pt}{2.8ex}}
 \newcommand\B{\rule[-1.4ex]{0pt}{0pt}}
  \begin{center}
    \begin{tabular}{llll}
      \hline
      \hline
       \T\B 
       $\boxed{n\rightarrow N\pi}$ \qquad\qquad& 
Total \qquad\qquad& Valence-sea \qquad\qquad& Sea-sea   \\
      \hline
      $n\rightarrow n\pi^0$  \qquad\qquad&\T $0$ \qquad\qquad& 
$2 q_u q_{\bar{u}}\,\chi^2_{K^+\Sigma^-} + 
     2 q_d q_{\bar{d}}\,(\chi^2_{K^0\Sigma^0}+\chi^2_{K^0\Lambda})$ \qquad\qquad& 
$q_{\bar{u}}^2\,\chi^2_{K^+\Sigma^-} + 
     q_{\bar{d}}^2\,(\chi^2_{K^0\Sigma^0}+\chi^2_{K^0\Lambda})$ \\
      $n\rightarrow p\pi^-$  \qquad\qquad&\T $2 (D+F)^2$ \qquad\qquad& 
$2 q_d q_{\bar{u}}\,(\chi^2_{K^0\Sigma^0}+\chi^2_{K^0\Lambda})$ 
    \qquad\qquad& $q_{\bar{u}}^2\,(\chi^2_{K^0\Sigma^0}+\chi^2_{K^0\Lambda})$\B\\
      $n\rightarrow n^-\pi^+$  \qquad\qquad&\T $0$ \qquad\qquad& 
$2 q_u q_{\bar{d}}\,\chi^2_{K^+\Sigma^-}$ 
    \qquad\qquad& $q_{\bar{d}}^2\,\chi^2_{K^+\Sigma^-}$\B\\
      \hline
\T\B 
$\boxed{n\rightarrow \Sigma K}$ \qquad\qquad& 
 \qquad\qquad&  \qquad\qquad&    \\
      \hline
      $n\rightarrow (\Sigma^0,\, \Lambda)\, K^0$  \qquad\qquad&\T $0$ \qquad\qquad& 
$2 q_d q_{\bar{s}}\,(\chi^2_{K^0\Sigma^0}+\chi^2_{K^0\Lambda})$ 
    \qquad\qquad& $q_{\bar{s}}^2\,(\chi^2_{K^0\Sigma^0}+\chi^2_{K^0\Lambda})$\\
      $n\rightarrow \Sigma^- K^+$  \qquad\qquad&\T $2 (D-F)^2$ \qquad\qquad& 
$2 q_u q_{\bar{s}}\,\chi^2_{K^+\Sigma^-}$ \qquad\qquad& 
$q_{\bar{s}}^2\,\chi^2_{K^+\Sigma^-}$ \B\\
      \hline
\T\B $\boxed{n\rightarrow \Delta\pi}$ \qquad\qquad& 
 \qquad\qquad&  \qquad\qquad&    \\
      \hline 
 $n\rightarrow \Delta^0\pi^0$  \qquad\qquad&\T $0$ \qquad\qquad& 
$2 q_u q_{\bar{u}}\,\chi_{K^+\Sigma^{*-}}^2 + 
2 q_d q_{\bar{d}}\,\chi_{K^0\Sigma^{*0}}^2$ \qquad\qquad& 
$q_{\bar{u}}^2\,\chi_{K^+\Sigma^{*-}}^2 + q_{\bar{d}}^2\,\chi_{K^0\Sigma^{*0}}^2$ \\
 $n\rightarrow \Delta^+\pi^-$\qquad\qquad&\T$\frac{4}{9}\ca{C}^2$
\qquad\qquad& 
$2 q_d q_{\bar{u}}\,\chi_{K^0\Sigma^{*0}}^2$ \qquad\qquad& 
$q_{\bar{u}}^2\,\chi_{K^0\Sigma^{*0}}^2$ \\
 $n\rightarrow\Delta^-\pi^+$\qquad\qquad&\T $\frac{4}{3}\ca{C}^2$
\qquad\qquad& 
$2 q_u q_{\bar{d}}\,\chi_{K^+\Sigma^{*-}}^2$ \qquad\qquad& 
$q_{\bar{d}}^2\,\chi_{K^+\Sigma^{*-}}^2$ 
\B\\
\hline
\T\B 
$\boxed{n\rightarrow \Sigma^* K}$ \qquad\qquad& 
\qquad\qquad&  \qquad\qquad&   \\
      \hline
      $n\rightarrow \Sigma^{*0} K^0$  \qquad\qquad&\T $0$ \qquad\qquad& 
$2 q_d q_{\bar{s}}\,\chi^2_{K^0\Sigma^{*0}}$ 
    \qquad\qquad& $q_{\bar{s}}^2\,\chi^2_{K^0\Sigma^{*0}}$\\
      $n\rightarrow \Sigma^{*-} K^+$  \qquad\qquad&\T $\frac{4}{9}\ca{C}^2$ \qquad\qquad& 
$2 q_u q_{\bar{s}}\,\chi^2_{K^+\Sigma^{*-}}$ \qquad\qquad& 
$q_{\bar{s}}^2\,\chi^2_{K^+\Sigma^{*-}}$ \B\\
      \hline
    \end{tabular}
  \end{center}
\vspace{-6pt}
  \label{table:nNpi}
\end{table*}

By repeating the above procedure for diagram~4$(c)$, one finds 
\begin{align}
\label{eqn:nc}
\chi_{nn\pi^0}^{diag (c)} &\propto (q_d^2 + 2 q_d q_{\bar{d}} +
q_{\bar{d}}^2) \, ,
\\ 
&\propto (q_d + q_{\bar{d}})^2  = 0 \, ,
\end{align} 
and
\begin{align}
\chi_{nn\pi^0}^{diag (c)} &\propto  \chi^2_{K^0\Sigma^0}+\chi^2_{K^0\Lambda}\nonumber \\
&\propto (D-F)^2 + \frac{1}{3}(D+3F)^2.
\label{eqn:gn8}
\end{align} 
As a result, one can now identify which components of the $n\rightarrow
n \pi^0$ channel have a disconnected sea-quark loop in the quark
flow.  Monitoring the quark charges that couple to the photons enables
one to identify the different valence-sea and sea-sea quark
sectors. Thus, the valence-sea and sea-sea contributions can be
calculated explicitly, as above.  Knowing the total coefficient from
standard $\chi$PT, the remainder represents the valence-valence
contribution including the connected quark flow of diagram 4$(a)$.
One may also apply this procedure to the $n\rightarrow p \pi^-$
channel.

In order to obtain the total partially quenched result for
$n\rightarrow N \pi$, one must also consider the unphysical process
$n\rightarrow n^- \pi^+$.  This process does not occur in full QCD
since the propagation of negatively charged $ddd$ neutronlike states
violates the Pauli exclusion principle.  This is realized in full QCD
by a cancellation of the two quark-flow diagrams associated with
$n\rightarrow n^- \pi^+$.  These diagrams are obtained from
Figs.~\ref{fig:nnpi0} (a) and ~\ref{fig:nnpi0} (c) with the change of the
valence flavour labels $d$, $d$, $u$ to $u$, $d$, $d$.  While the sum
of these two diagrams vanishes, each one participates in the leading-order 
nonanalytic coefficients of the magnetic polarizability.  An
omission of photon couplings to the disconnected sea-quark loop in the
lattice QCD simulations allows a nontrivial contribution.  The
cancellation no longer takes place, and this must be
taken into account when fitting the lattice results.

The diagrammatic procedure may be repeated for $n\rightarrow
\Delta\pi$.  A summary of the contributions in different channels is
shown in Table~\ref{table:nNpi} for both octet and decuplet
transitions.  In summary, the modifications to the loop integrals of
Eqs.~(\ref{eqn:loopa}) and (\ref{eqn:loopb}) due to partial quenching
in the lattice QCD simulations are
\begin{align}
\label{eqn:pQN}
\chi_N \to \chi_N^{pQ} &= 2 g_A^2 -(D-F)^2 - \frac{7}{27}(D+3F)^2,\\
\chi_\Delta \to \chi_\Delta^{pQ}&=\frac{16}{9}\ca{C}^2 -
\frac{2}{9}\ca{C}^2 \, .
\label{eqn:pQD}
\end{align} 
Note that these coefficients are consistent with those derived from the
graded-symmetry approach (see Table I of Ref.~\cite{Detmold:2006vu}).

Because the lattice simulations incorporate $2+1$ flavours, the kaon
loops also require consideration.  Given the international attention
devoted to learning the strangeness contribution to the magnetic
moment of the nucleon, it is fascinating to perform a similar
calculation \cite{Leinweber:2004tc} for the magnetic polarizability.  To achieve
this, additional loop integrals are considered with the same form as
Eq.~(\ref{eqn:loopb}) but with the pion replaced by the kaon and the
mass splitting associated with the increased mass of the hyperons in
the intermediate states.
  
Specifically, the symbol $\Delta$ in Eq.~(\ref{eqn:loopb}) represents
the mass splitting between the nucleon and either the $\Sigma$ or the
$\Sigma^*$ baryon; taking the experimental charge-state averages 
$m_\Sigma = 1.189$ GeV and $m_{\Sigma^*} = 1.383$ GeV.  The kaon mass
is related to the pion mass via
\begin{equation}
m_K^2 = m_{K,\ro{phys}}^2 + \frac{1}{2}(m_\pi^2-m_{\pi,\ro{phys}}^2). 
\end{equation}
The coefficient for the partially strange quark contribution to the
neutron magnetic moment is obtained with the modification of
$\chi_\Delta$ in Eq.~(\ref{eqn:loopb})
\begin{equation}
\chi_\Delta \to \chi_{K\Sigma}^{pQ} = 2(D - F)^2 - (D - F)^2 + \frac{1}{27}(D+3F)^2 \, , 
\label{eqn:strangeKS}
\end{equation} 
and
\begin{equation}
\chi_\Delta \to \chi_{K\Sigma*}^{pQ} = \frac{4}{9} \ca{C}^2 \, - \frac{2}{9} \ca{C}^2 \, , 
\label{eqn:strangeKSstar}
\end{equation} 
where the first term on the right-hand side of these equations is the
full QCD contribution.

The strange sea-quark-loop contribution to the nucleon magnetic
polarizabilities can be obtained via
\begin{equation}
\chi_\Delta \to \chi_{K\Sigma}^{s} = \frac{1}{3}(D - F)^2 + \frac{1}{27}(D+3F)^2
\label{eqn:strangeOnly}
\end{equation} 
and
\begin{equation}
\chi_\Delta \to \chi_{K\Sigma^*}^{s} = \frac{2}{27} \ca{C}^2 \, ,
\label{eqn:strangeOnlyStar}
\end{equation} 
where the square of the strange-quark charge factor of $1/9$ is included in the coefficients.

In the next section, The lattice results are treated with
partially quenched $\chi$EFT, i.e. using the coefficients in
Eqs.~(\ref{eqn:pQN}), (\ref{eqn:pQD}), (\ref{eqn:strangeKS}) and
(\ref{eqn:strangeKSstar}).  The full QCD coefficients can be recovered
by keeping only the first term on the right-hand side of these
equations.  The effect of unquenching the missing light-quark
disconnected loop contributions is investigated first, followed by a
complete restoration of sea-quark loop contributions through the
inclusion of strange-quark-loop contributions in the kaon dressings.

\section{Chiral extrapolation}
\label{sec:extrap}

\begin{figure*}
\begin{minipage}[t]{0.4\linewidth}
\includegraphics[height=1.0\hsize,angle=90]{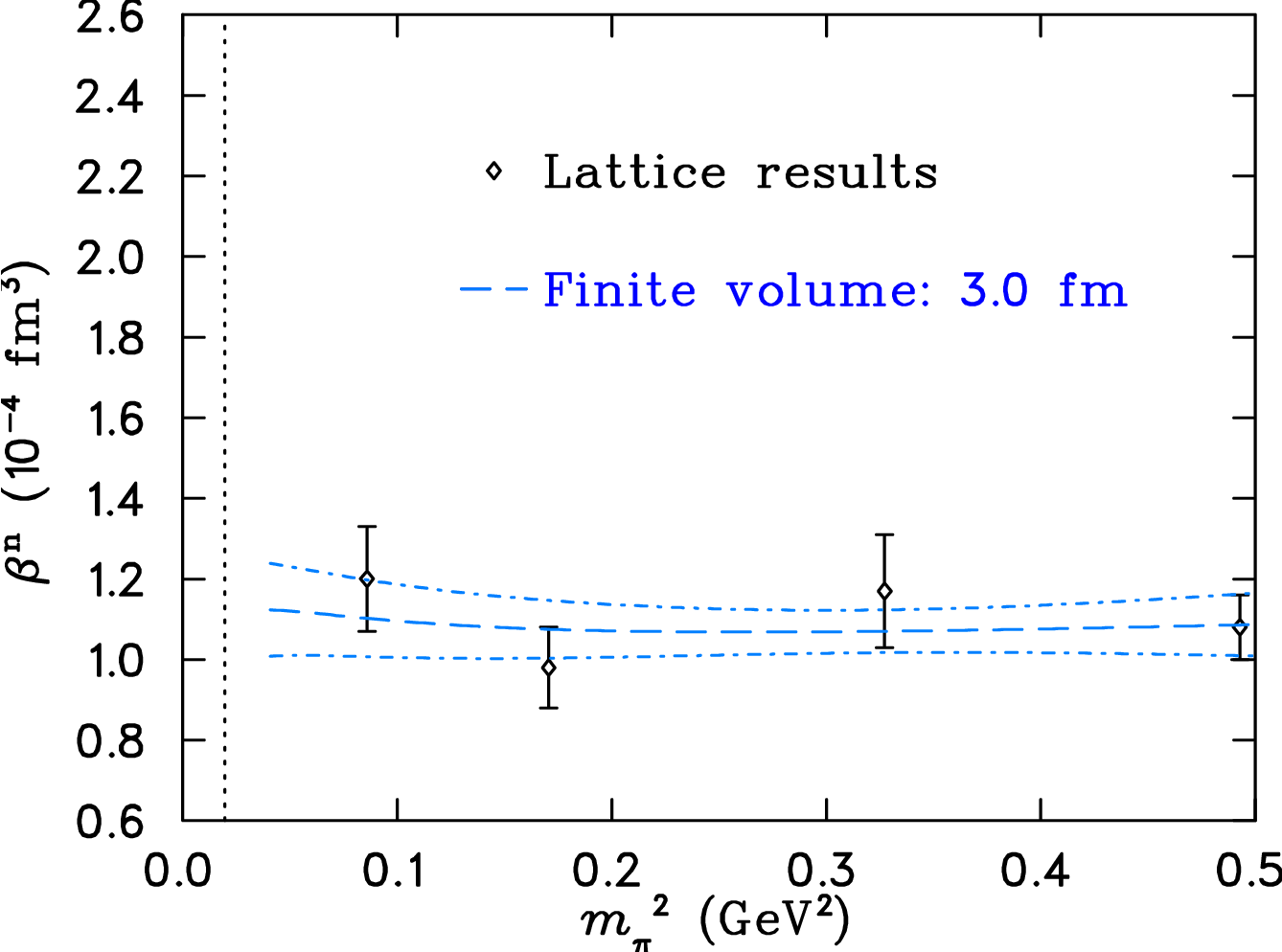}
\vspace{-5mm}
\caption{\footnotesize{(color online). Extrapolation of the magnetic
    polarizability of the neutron, $\beta^n$, at spatial length $L =
    3.0$ fm. 
The lattice points satisfy $m_\pi L > 3$.
The dot-dashed curves indicate the error bar
    associated with the fit. 
    The vertical dotted line indicates the physical point.  }}
\label{fig:extrapfin}
\vspace{5mm}
\includegraphics[height=1.0\hsize,angle=90]{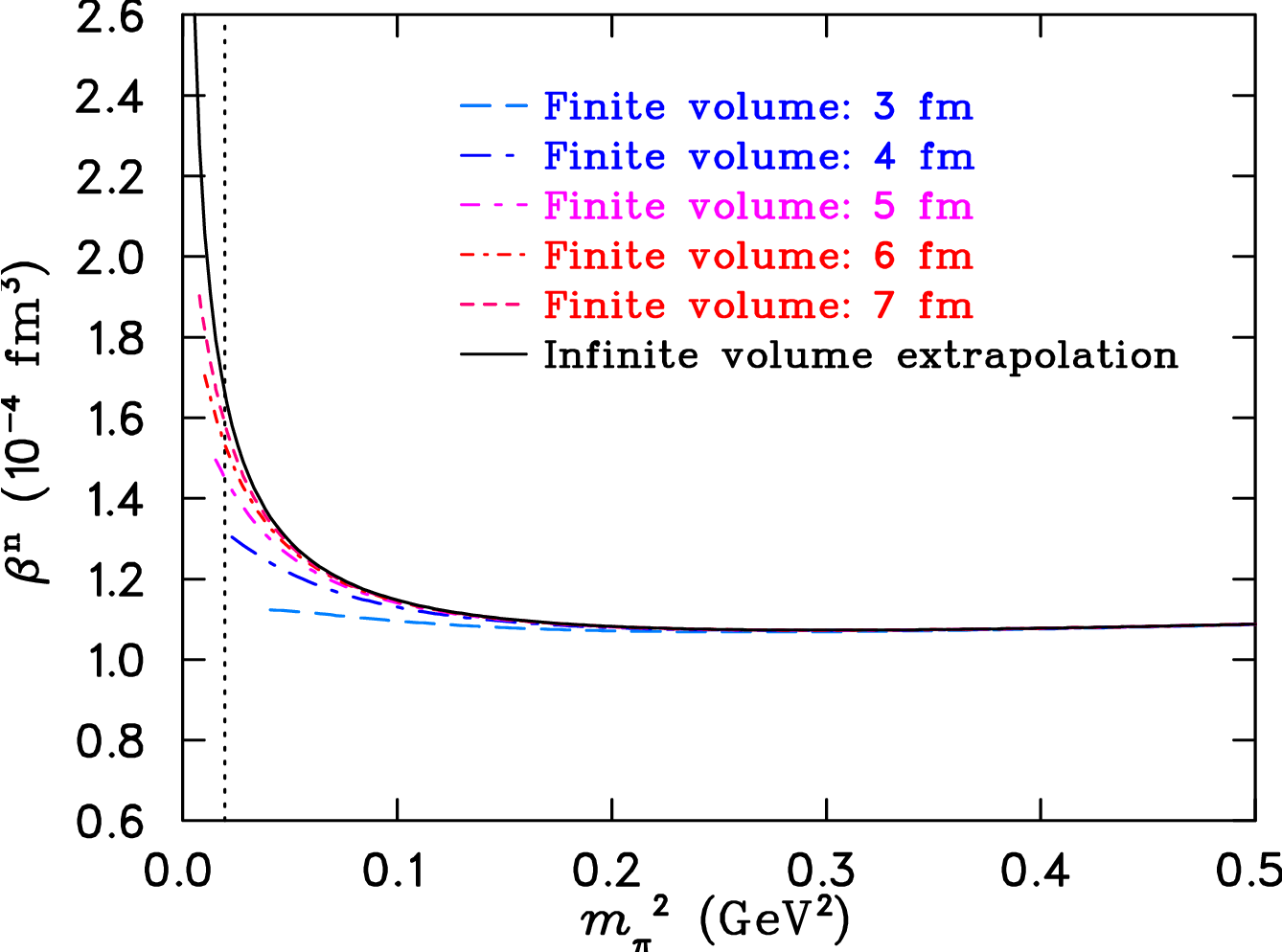}
\vspace{-5mm}
\caption{\footnotesize{(color online). Extrapolation of the magnetic
    polarizability of the neutron, $\beta^n$, for a variety of spatial
    lattice volumes, and the infinite volume limit.  }}
\label{fig:extrapmultifin}
%
%
\vspace{5mm}
\includegraphics[height=1.0\hsize,angle=90]{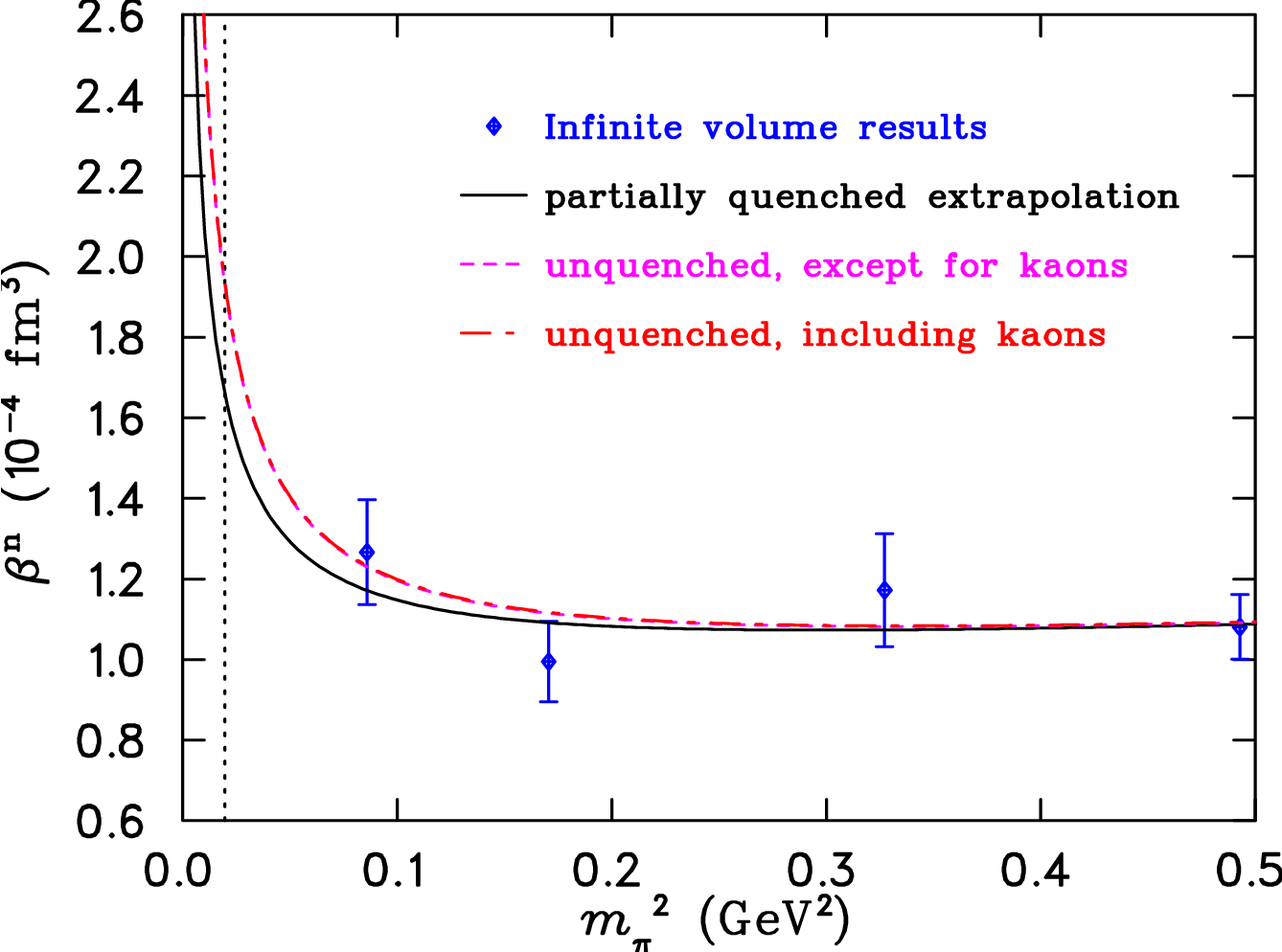}
\vspace{-5mm}
\caption{\footnotesize{(color online). A comparison of the
    extrapolations of the magnetic polarizability of the neutron,
    $\beta^n$, upon including the contributions of photon couplings to
    the disconnected $u$, $d$ and $s$ quark loops which were omitted
    in the lattice QCD simulations.  
%
}}
\label{fig:extrapunQ}
\end{minipage}
\hspace{12mm}
\begin{minipage}[t]{0.4\linewidth}
\includegraphics[height=1.0\hsize,angle=90]{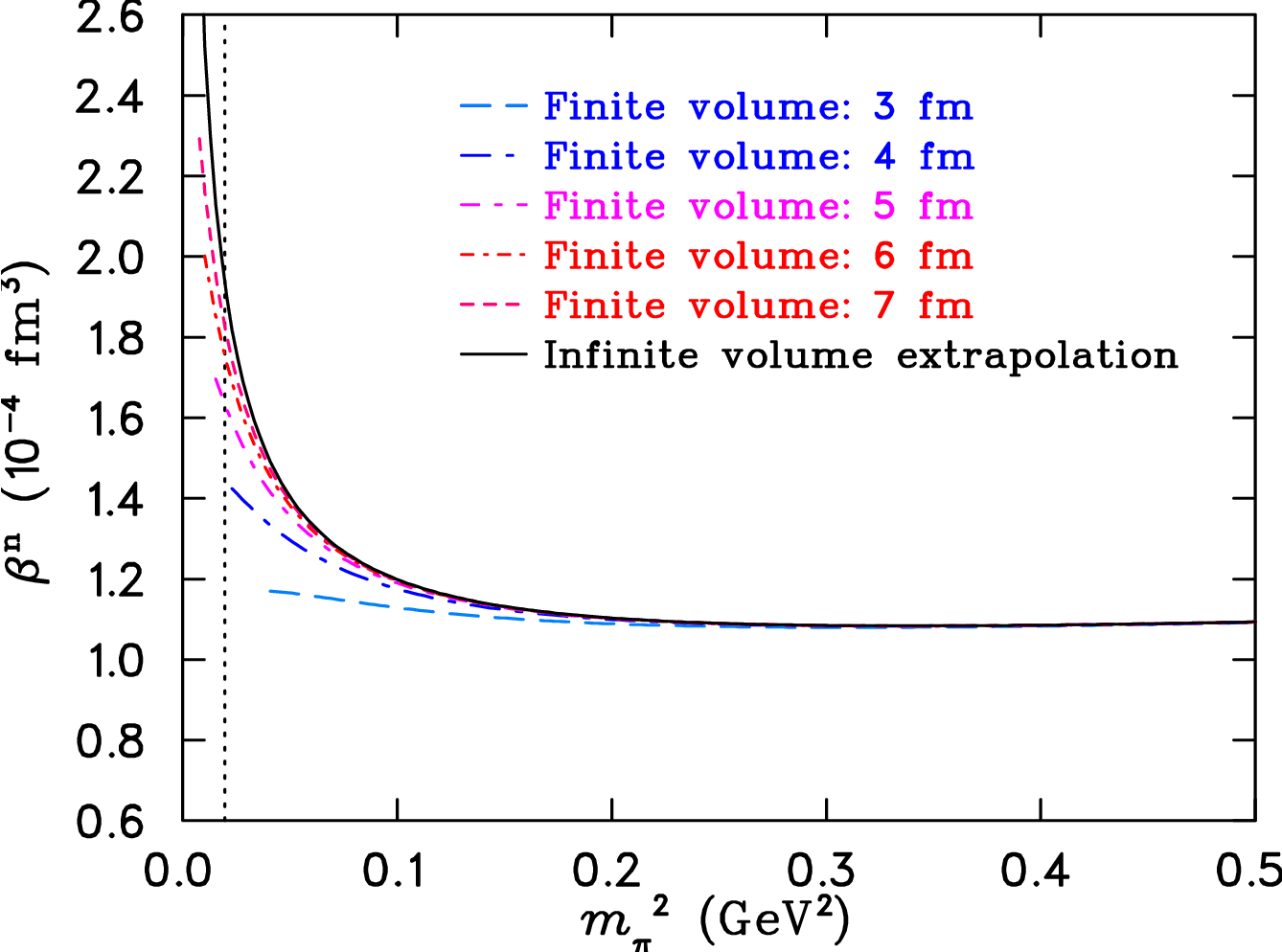}
\vspace{-11pt}
\caption{\footnotesize{(color online). Extrapolation of the unquenched
    value of $\beta^n$ for a variety of spatial lattice volumes, and
    the infinite volume case, correcting for partial quenching effects.  
These results provide a benchmark to guide
    the interpretation of future lattice QCD simulations including
    background field effects in the disconnected sea-quark loop
    sector.  }}\label{fig:extrapmultifinunQ}
\vspace{5mm}
\includegraphics[height=1.0\hsize,angle=90]{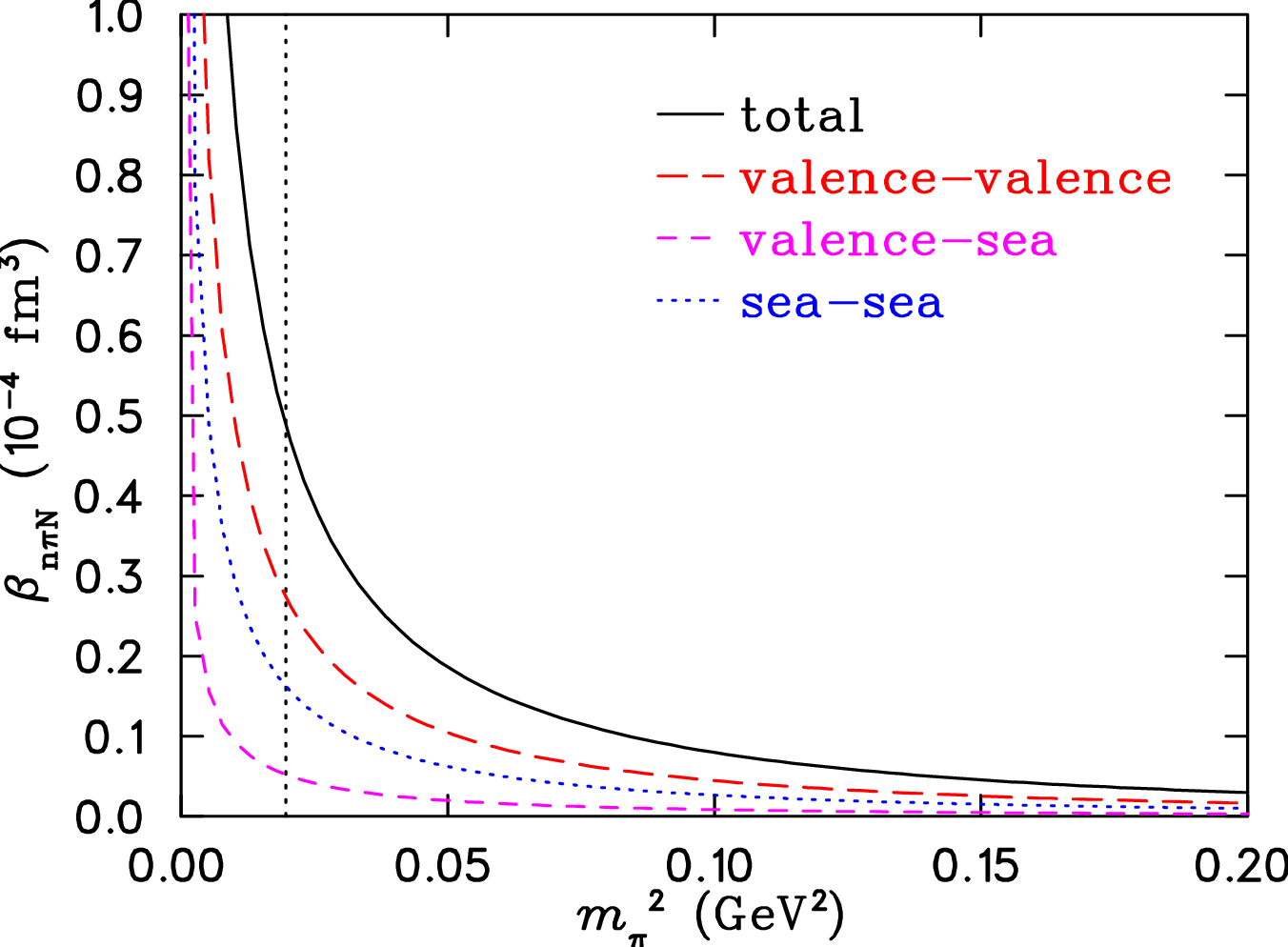}
\vspace{-11pt}
\caption{\footnotesize{(color online). The contributions from separate
    photon-quark coupling scenarios to the leading-order octet loop
    integral of Eq.~(\ref{eqn:loopa}).  In the valence-sea case where
    one photon couples to a valence quark and the other couples to a
    sea quark, there is a large positive contribution from
    $n\rightarrow K^0 \Lambda$, and the overall valence-sea result is
    positive.  }}
\label{fig:sectorspiN}
\vspace{5mm}
\includegraphics[height=1.0\hsize,angle=90]{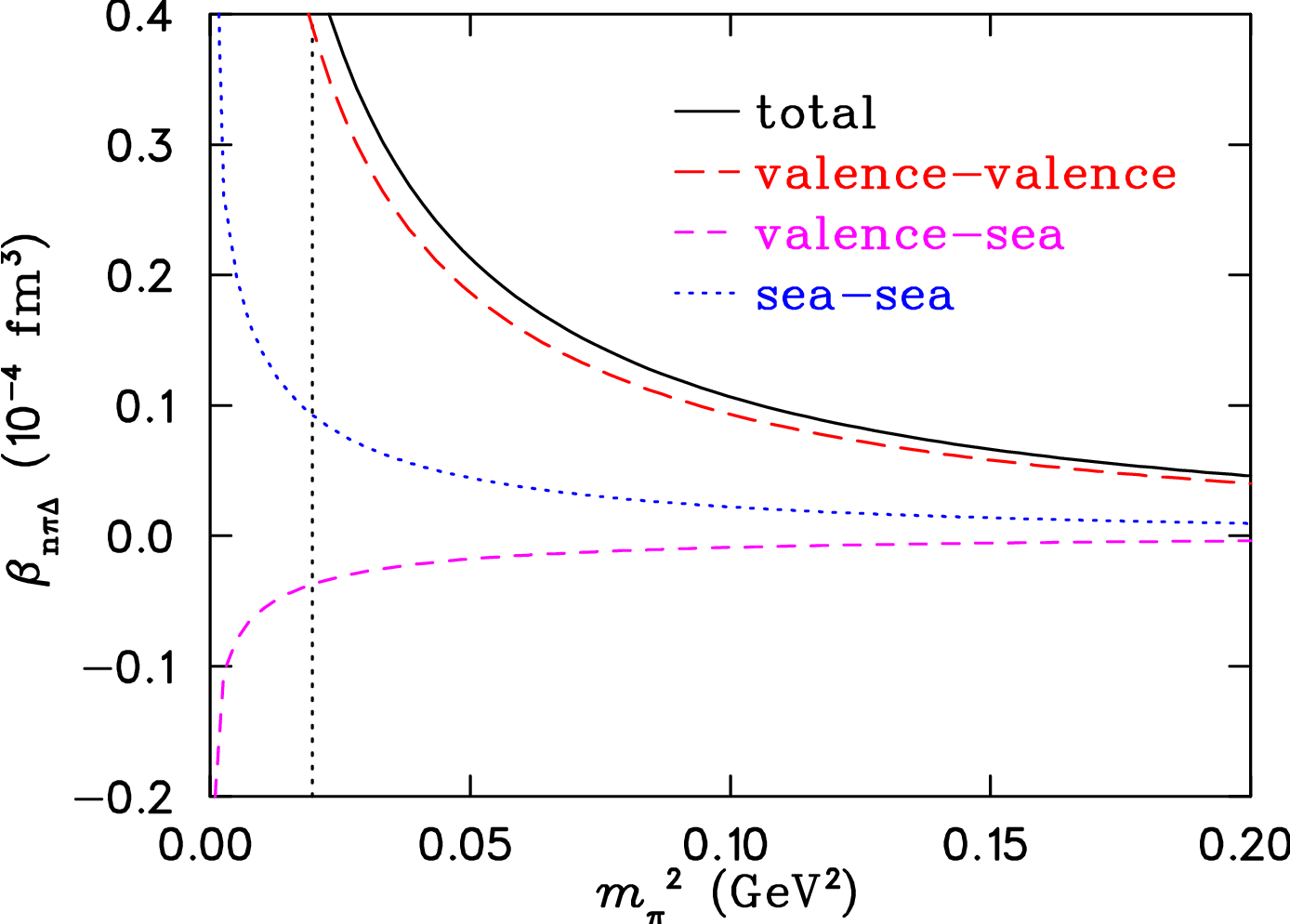}
\vspace{-11pt}
\caption{\footnotesize{(color online). The contributions from separate
    photon-quark coupling scenarios to the next-to-leading-order
    decuplet loop integral of Eq.~(\ref{eqn:loopb}).  In the
    valence-sea case, the negative contribution from $n\rightarrow
    K^+\Sigma^*$ dominates, and the overall valence-sea result is
    negative.  }}
\label{fig:sectorspiD}
\end{minipage}
\end{figure*}

The finite-volume correction and chiral extrapolation of the magnetic
polarizability of the neutron can now be performed with the lattice
simulation results of Ref.~\cite{Primer:2013pva}.  The $2+1$ flavour
simulation results are illustrated in Fig.~\ref{fig:extrapfin} with a
partially quenched finite-volume extrapolation suitable for these
lattice results.  

The extrapolation includes a term linear in $m_\pi^2$ with coefficient
$a_2$ determined by a fit to the lattice results.  The fit value of
$a_2$ is small, at $6.5\times 10^{-7}\,\ro{fm}^5$.  
In illustrating the extrapolation curve, the results for $m_\pi L< 3$ are
 not shown, and it is noted 
that the magnetic polarizability at the physical point cannot be
reached with a (3 fm)${}^3$ volume. 
Alternatively, one could apply the more conservative constraint 
of $m_\pi L > 4$ without changing the shapes of the extrapolation curves, 
as all lattice points used satisfy $m_\pi L > 4.45$. In Figs.~\ref{fig:extrapfin}, \ref{fig:extrapmultifin} and \ref{fig:extrapmultifinunQ}, 
the preference is to illustrate the results over a wider range.

Since the Sommer scale has been selected, the lattice volume varies
slightly across the four lattice points available.  However, the
finite-volume corrections for large pion-mass values are relatively
small, as illustrated in an example extrapolation for spatial length $3.0$
fm, which corresponds to the volume of the lightest point at $m_\pi =
293$ MeV.  Differences between the results for the varying volume and
those for a $3.0$ fm box at the higher pion-mass points are very small
and cannot be seen in Fig.~\ref{fig:extrapfin}.

Chiral extrapolations for a range of finite volumes, and the infinite
volume, are shown in Fig.~\ref{fig:extrapmultifin}. This highlights
the manner in which the discretization of momenta to only those
available on the finite volume significantly suppresses the chiral
dressings of the neutron.  The anticipated chiral curvature is
significantly reduced on smaller lattice volumes.  As precision
lattice results become available, these curves can provide an
important benchmark in understanding the volume dependence of those
results.  


For volumes of $(4\ \mbox{fm})^3$ and larger, finite-volume
corrections are significant only for $m_\pi < 300$ MeV but they grow
quickly in the chiral regime.  Consequently, box sizes as large as $7$
fm are required to obtain an extrapolation within $5\%$ of the
infinite-volume value at the physical point.

The effect of unquenching the disconnected loops, by replacing the
meson dressings of partially quenched QCD with those of full QCD, is
shown in Fig.~\ref{fig:extrapunQ} at infinite volume.  One constrains
the analytic terms, $a_0 + a_2\, m_\pi^2$, of the chiral expansion
by fitting the partially quenched chiral expansion to
the partially quenched lattice simulation results corrected to
infinite volume.  The black curve of Fig.~\ref{fig:extrapunQ}
illustrates this fit. 

With the regulator parameter fixed to $\Lambda = 0.8$ GeV the analytic
terms model the invariant core contribution
\cite{Young:2002cj,Leinweber:2004tc,Leinweber:2006ug,Wang:1900ta,Wang:2008vb,Wang:2013cfp}
to the polarizability.  One can then correct the meson-cloud
contribution by adding the valence-sea and sea-sea-loop integrals.
Figure \ref{fig:extrapunQ} illustrates the important effect of
unquenching the light $u$ and $d$ sea-quark sector contributions to
the magnetic polarizability with the magenta dashed curve.  The final
effect of fully unquenching the results by also unquenching the
strange sector through the addition of kaon loops is illustrated by
the red dot-dashed curve.

The unquenched theory displays a significant increase in the 
strength of the chiral loop integrals with the value at the physical
point sitting higher than in the partially quenched case.  At the
physical point, the magnetic polarizability is only $1.66 \times
10^{-4}\ \mbox{fm}^3$, whereas the full theory provides $1.93 \times
10^{-4}\ \mbox{fm}^3$, a $16\%$ correction.

In contrast, unquenching the kaon loops has a very tiny effect, with a
percentage shift of approximately $0.16\%$.  This is significantly
smaller than the effect of the strange-quark contributions to the
proton magnetic moment \cite{Leinweber:2004tc} of $0.55\%$.
%
Isolating the strange-quark sea-sea contribution via 
Eqs.~(\ref{eqn:strangeOnly}) and (\ref{eqn:strangeOnlyStar}) 
provides a more closely related comparison of strange-sea-quark-loop 
contributions.  In this case the strange-sea-quark-loop contribution 
to the magnetic polarizability is $0.0023\times 10^{-4}\,\ro{fm}^3$, 
a $0.12\%$ contribution.

A comparison of multiple finite-volume and infinite-volume
extrapolations for full QCD is shown in
Fig~\ref{fig:extrapmultifinunQ}, correcting for partial-quenching effects.  
These results thus provide a benchmark to
guide the interpretation of future lattice QCD simulations including
background field effects in the disconnected sea-quark-loop sector.

The breakdown of the loop integral contributions into the
valence-valence, valence-sea and sea-sea contributions is illustrated
in Fig.~\ref{fig:sectorspiN} for the $\pi N$ sector of
Eq.~(\ref{eqn:loopa}) and Fig.~\ref{fig:sectorspiD} for the $\pi
\Delta$ sector of Eq.~(\ref{eqn:loopb}).  The difference in the sign
of the valence-sea contribution between the two plots is noteworthy,
as it highlights an important difference between the octet and
decuplet processes.  The quark-flow diagrams corresponding to a
neutral intermediate ($n$ or $\Delta^0$) are the same, and the
valence-sea contribution from each is negative due to the opposite
charges of the $q \overline q$ pairs.  However, in the case of the
octet, the large contributions from $K^0 \Lambda$-type coupling for
the disconnected $u$-quark loop in the $n\rightarrow p \pi^-$ channel
dominate over the similar $d$-quark loop in the neutral channel.  The
$\overline u$- and $d$-quark charges multiply positively, and the
valence-sea contribution is positive.  In the decuplet, there is no
equivalent large coupling, and the neutral channel,
$n\rightarrow\Delta^0\pi^0$, dominates, causing the valence-sea
contribution to be negative.

\begin{figure*}[p]
\includegraphics[height=0.7\hsize,angle=90]{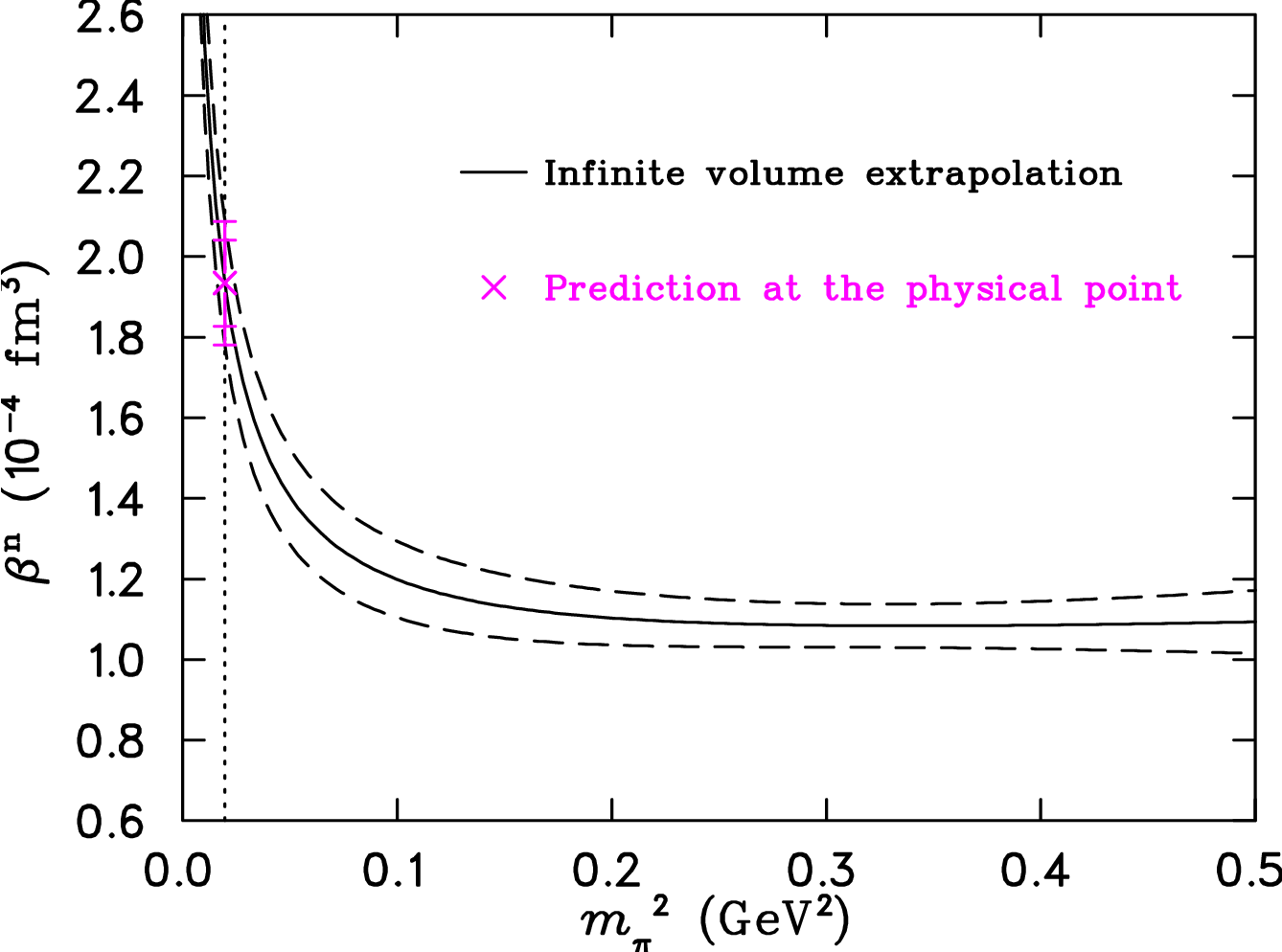}
\vspace{-10pt}
\caption{\footnotesize{(color online).  Our prediction for the pion
    mass dependence of the magnetic polarizability of the neutron,
    $\beta^n$, is illustrated by the solid curve with dashed curves
    illustrating combined statistical and systematic uncertainties.  At
    the physical point, the inner error bar represents the statistical
    uncertainty from the fit to the lattice data, and the outer error
    bar adds the systematic uncertainty from the meson-cloud parameter,
    $\Lambda$, in quadrature. }}
\label{fig:extrapinf}
\vspace{5mm}
\null\qquad\;
\includegraphics[height=0.66\hsize,angle=90]{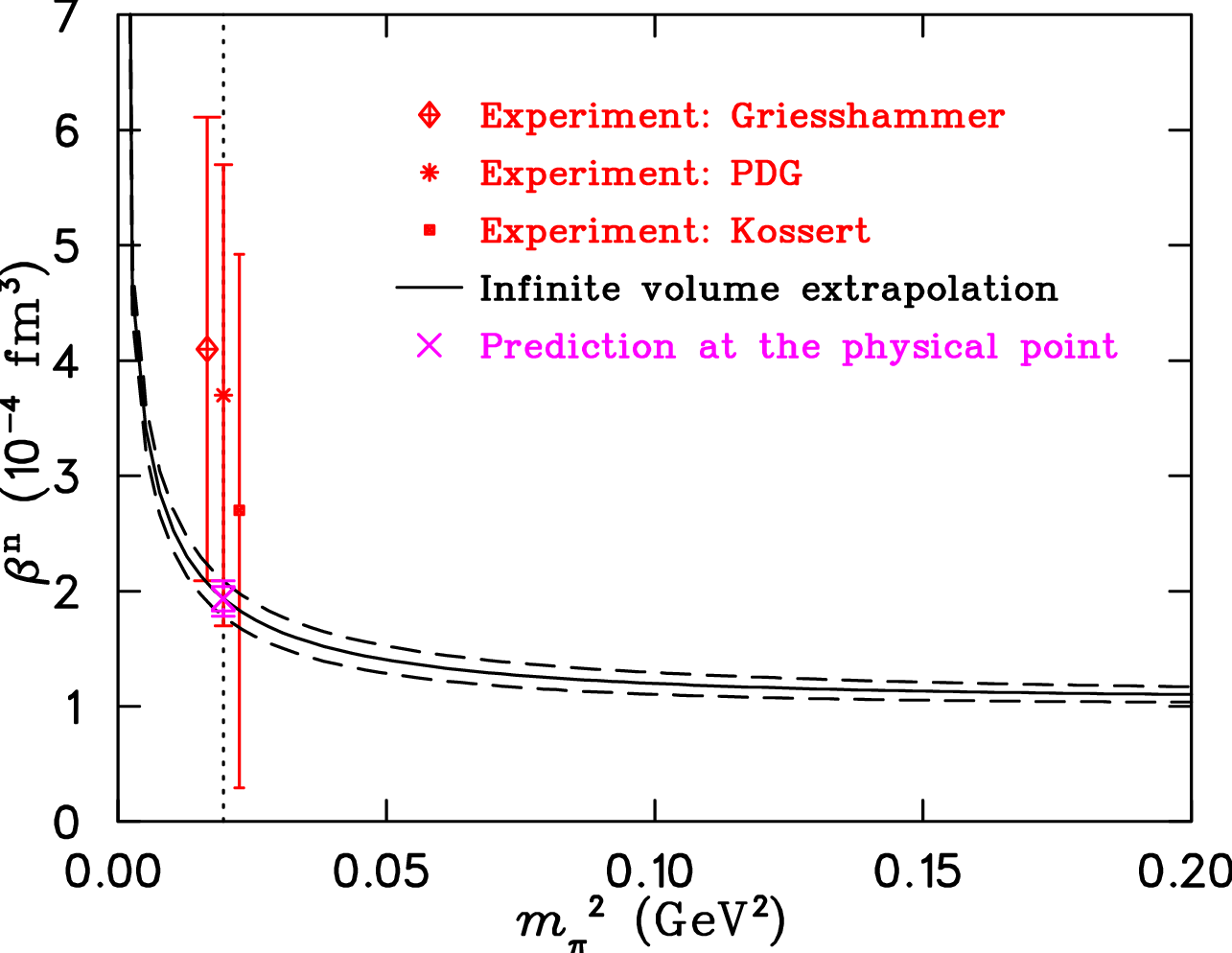}
\vspace{-10pt}
\caption{\footnotesize{(color online).  The magnetic polarizability of
    the neutron, $\beta^n$, is compared with experimental results.
    Uncertainties contain both statistical and systematic errors added
    in quadrature.  Experimental results from Grie{\ss}hammer
    \emph{et. al.}  \cite{Griesshammer:2012we}, the PDG
    \cite{Beringer:1900zz}, and Kossert \emph{et. al.}
    \cite{Kossert:2002jc,Kossert:2002ws} are offset for clarity.  }}
\label{fig:extrapinfexpt}
\end{figure*}

The final infinite-volume full-QCD prediction for the magnetic
polarizability of the neutron is shown in Fig.~\ref{fig:extrapinf},
with a value of $\beta^n = 1.93(11)(8)\times 10^{-4}\,\ro{fm}^3$ at
the physical point.  The quoted uncertainties represent both the
statistical error from constraining the fit parameters to lattice QCD
results, and the systematic uncertainty from variation of $\Lambda$
over the range $0.7 \le \Lambda \le 0.9$.  In the plot, the inner
error bar represents the statistical uncertainty from the fit only,
and the outer error bar includes the systematic uncertainty from the
regulator parameter $\Lambda$ added in quadrature. 
Since the lattice results are obtained using a single lattice-spacing, 
it is not possible to quantify an uncertainty associated with 
taking the continuum limit. However, the lattice calculations are performed 
using a nonperturbatively improved clover-fermion action, and 
it is therefore anticipated that the $\ca{O}(a^2)$ corrections are small 
relative to the uncertainties already addressed. 

A comparison between our result and the experimental data is shown in
Fig.~\ref{fig:extrapinfexpt}.  In addition to the Particle Data Group
value \cite{Beringer:1900zz}, analyses of elastic photon-deuteron
scattering experiments by Grie{\ss}hammer \emph{et. al.}
\cite{Griesshammer:2012we} and Kossert \emph{et. al.}
\cite{Kossert:2002jc,Kossert:2002ws} are included in the plot.  For
clarity of comparison, an $m_\pi^2$-axis offset is introduced among
the experimental points.  Our result is in good agreement with all
three experimental measurements and presents an interesting challenge
for greater precision in the experimental measurement.  Such progress
would similarly drive further progress in lattice QCD simulations and
chiral effective field theory.

\section{Conclusion}

Dynamical lattice QCD simulation results for the magnetic polarizability of the
neutron have only recently become available \cite{Primer:2013pva}.
The results are obtained at finite volume and only the quarks carrying
the quantum numbers of the hadron experience the background field.
The dynamical fermion loops of the QCD simulation are blind to the
external field.  As such, it is timely to investigate the physics
required to relate these new partially quenched simulation results to
experiment.

Heavy baryon chiral effective field theory provides a framework
in which to perform this investigation.  Methods to correct for the
finite volume of the spatial lattice volume are well established and
employed herein.  Techniques to unquench the sea-quark-loop
contributions are also well established
\cite{Young:2002cj,Leinweber:2004tc,Leinweber:2006ug,Wang:2008vb} but
their application to the magnetic polarizability herein is novel.

This work has made definitive progress in providing a theoretical
prediction for the magnetic polarizability of the neutron.  We find
$\beta^n = 1.93(11)^{\ro{stat}}(8)^{\ro{sys}} \times
10^{-4}\,\ro{fm}^3$.  The prediction is founded on first-principles
lattice QCD simulations and incorporates effective field theory
techniques to correct for the finite volume of the lattice, account
for the disconnected sea-quark-loop contributions and connect to the
light quark masses of Nature.  The result agrees with current
experimental estimates and presents an interesting challenge for
greater precision in the experimental measurement.

In performing the chiral extrapolations the finite-volume effects were
quantified for a range of spatial lattice volumes relevant to current
and future lattice simulations.  Both partially quenched and full QCD
results were addressed in the finite-volume analysis.  It was found
that lattices of approximately $7$ fm on a side are required to obtain
the magnetic polarizability of the neutron to within $5\%$ of the
infinite-volume value at the physical pion mass.  These finite-volume
studies provide a benchmark for future lattice QCD calculations and a
guide to the interpretation of the results.

Unquenching the disconnected-loop contributions provides a
significant increase in the chiral curvature  of the magnetic
polarizability and a significantly larger prediction at
the physical point.  Unquenching the $u$, $d$ and $s$ disconnected
loop contributions resulted in a $16\%$ increase in the infinite-volume 
prediction. 
The contribution from kaon loops is negligibly small, at $0.16\%$.
This is smaller than the $0.55\%$ effect associated with strange-quark
contributions to the proton's magnetic moment \cite{Leinweber:2004tc}.

A more precise experimental measurement of the magnetic
polarizabilities of the nucleon is clearly warranted.  Similarly,
further investment in lattice QCD investigations is of value. In 
the case of lattice QCD, 
the difficult problem of Landau-level contributions to the 
correlation functions is of interest, as is the need to directly
incorporate sea-quark-loop effects.  Finally, higher-order terms of
the chiral expansion are valuable in evaluating the convergence
of the effective field theory.


\begin{acknowledgments}
J.M.M.H. is thankful to Judith McGovern and Thomas Primer for helpful
discussions.  This research is supported by the Australian Research
Council through the ARC Centre of Excellence for Particle Physics at
the Terascale, and through Grants No. DP120104627 (D.B.L.),
No. DP110101265 (D.B.L. and R.D.Y.) and No. FT120100821 (R.D.Y.).
\end{acknowledgments}

\bibliographystyle{apsrev} 
\bibliography{refs}

\end{document}